\documentclass[seceq]{ptptex}
\usepackage{wrapft}
\usepackage{bm}

\title{
Equivalence Classes of Boundary Conditions in Gauge Theory on $Z_3$ Orbifold
}

\author{
Yoshiharu \textsc{Kawamura},\footnote{E-mail: haru@azusa.shinshu-u.ac.jp}
Teppei \textsc{Kinami} 
and Takashi \textsc{Miura}
}

\inst{
Department of Physics, Shinshu University, Matsumoto 390-8621, Japan
}

\recdate{
August 21, 2008}

\abst{
We study equivalence classes of boundary conditions in a gauge theory on the orbifold $T^2/Z_3$.
Orbifold conditions and those gauge transformation properties are given
and the gauge equivalence is understood by the Hosotani mechanism.
Mode expansions are carried out for six-dimensional $Z_3$ singlet fields and a $Z_3$ triplet field,
and the one-loop effective potential for Wilson line phases is calculated.
}

\begin{document}

\maketitle

\section{Introduction}

The boundary conditions (BCs) to be imposed on the fields in the bulk 
are classified into the equivalence classes using the gauge invariance, in higher-dimensional gauge theories.
Several sets of BCs belong to the same equivalence class and describe the same physics,
if they are related to gauge transformations.
Specifically, the symmetry of BCs is not necessarily the same as the physical symmetry.
The physical symmetry is determined by the Hosotani mechanism
after the rearrangement of gauge symmetry.\cite{H}

Grand unified theories on an orbifold have been attracted phenomenologically
since Higgs mass splitting was well realized by the orbifold breaking mechanism.\cite{Kawamura1,Hall1}$^,$\footnote{
In four-dimensional heterotic string models, extra colored Higgs are projected by the Wilson line mechanism.\cite{Higgs}}
Equivalence classes of BCs and dynamical gauge symmetry breaking were studied for gauge theories
on the orbifolds $S^1/Z_2$\cite{HHHK,HHK}$^,$\footnote{
See Ref. \citen{KLY} for the breakdown of gauge symmetry on $S^1/Z_2$ by the Hosotani mechanism.}
and $T^2/Z_2$.\cite{HN&T}
It is interesting to study equivalence classes of BCs and the Hosotani mechanism 
for gauge theories on other orbifolds and to construct a phenomenologically viable model based on them.
The $Z_3$ orbifold $T^2/Z_3$ is a candidate and
has been utilized in the search for the origin of three families\cite{threefamilies}
and the unification of gauge, Higgs and family.\cite{GLM&S}$^,$\footnote{
The six-dimensional extension of $Z_3$ orbifold was initially introduced into the construction of
four-dimensional heterotic string models.\cite{orbifold}}

In the present paper, we study equivalence classes of BCs in a gauge theory on $T^2/Z_3$.
Orbifold conditions and those gauge transformation properties are given
and the gauge equivalence is understood by the Hosotani mechanism.
Mode expansions are carried out for six-dimensional $Z_3$ singlet fields and a $Z_3$ triplet field,
and the one-loop effective potential for Wilson line phases is calculated.

In \S 2, general arguments are given for BCs in gauge theories on $T^2/Z_3$, 
and equivalence classes of BCs are defined by the invariance under the gauge transformation.
In \S 3, mode expansions on six-dimensional fields are given 
and the classification of BCs for the $SU(N)$ gauge group is carried out with the aid of equivalence relations.
The one-loop effective potential for Wilson line phases is calculated using an $SU(3)$ gauge theory.
Section 4 is devoted to conclusions.

\section{Orbifold conditions and equivalence classes}

\begin{wrapfigure}{l}{6.6cm}
\label{F1.Z3}
\unitlength 0.1in
\begin{picture}( 48.6400, 13.5000)( 13.0000,-18.5400)
%
\special{pn 13}%
\special{pa 2282 1704}%
\special{pa 3638 1698}%
\special{fp}%
\special{sh 1}%
\special{pa 3638 1698}%
\special{pa 3570 1678}%
\special{pa 3584 1698}%
\special{pa 3570 1718}%
\special{pa 3638 1698}%
\special{fp}%
%
\special{pn 13}%
\special{pa 2282 1698}%
\special{pa 1614 582}%
\special{fp}%
\special{sh 1}%
\special{pa 1614 582}%
\special{pa 1630 650}%
\special{pa 1640 628}%
\special{pa 1664 630}%
\special{pa 1614 582}%
\special{fp}%
%
\special{pn 13}%
\special{pa 3630 1698}%
\special{pa 2944 566}%
\special{fp}%
%
\special{pn 13}%
\special{pa 1612 574}%
\special{pa 2952 566}%
\special{fp}%
%
\special{pn 20}%
\special{sh 1}%
\special{ar 2282 1698 10 10 0  6.28318530717959E+0000}%
\special{sh 1}%
\special{ar 2282 1698 10 10 0  6.28318530717959E+0000}%
%
\special{pn 20}%
\special{sh 1}%
\special{ar 2268 948 10 10 0  6.28318530717959E+0000}%
\special{sh 1}%
\special{ar 2268 948 10 10 0  6.28318530717959E+0000}%
%
\special{pn 20}%
\special{sh 1}%
\special{ar 2930 1316 10 10 0  6.28318530717959E+0000}%
\special{sh 1}%
\special{ar 2930 1316 10 10 0  6.28318530717959E+0000}%
\put(20.3100,-18.7000){\makebox(0,0)[lb]{$z_0$}}%
\put(27.4600,-15.3100){\makebox(0,0)[lb]{$z_1$}}%
\put(21.2200,-11.3500){\makebox(0,0)[lb]{$z_2$}}%
\put(37.0000,-17.6000){\makebox(0,0)[lb]{$e_1$}}%
\put(14.3000,-5.6000){\makebox(0,0)[lb]{$e_2$}}%
%
\end{picture}%
\caption{Orbifold $T^2/Z_3$.}
\end{wrapfigure}
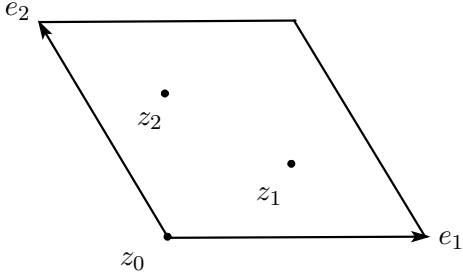
Let $x$ and $z$ be coordinates of $M^4$ and $T^2/Z_3$, respectively. 
$T^2$ is the two-dimensional torus whose basis vectors are 
$SU(3)$ root vectors, $e_1 = 1$ and $\displaystyle{e_2 = e^{2\pi i/3} \equiv \omega}$.\footnote{
We take the $SU(3)$ lattice as the unit lattice. 
On the estimation of physical quantities, we use physical sizes such as
$e_1 = 2 \pi R$ and $\displaystyle{e_2 = 2 \pi R \omega}$.}
On $T^2$, the point $z$ is identified by $z+n_1 e_1 +n_2 e_2$, 
where $n_1$ and $n_2$ are integers.
$T^2/Z_3$ is obtained by further identifying points on $T^2$ through a $Z_3$ rotation, i.e.,
$z$ is identified with $\theta z$ where $\theta^3 = 1$.
The resultant space is the area depicted in Fig.~1, which contains the information on $T^2$.

\subsection{Boundary conditions}

The fixed points $z_{\rm fp}$ on $T^2/Z_3$ are points that transform themselves
under the $Z_3$ transformation $z \to \theta z$ and satisfy
\begin{eqnarray}
z_{\rm fp} = \theta z_{\rm fp} + n e_1 + m e_2 ,
\label{zfp}
\end{eqnarray}
where $n$ and $m$ are integers that characterize fixed points.
There are three kinds of fixed points, namely,
\begin{eqnarray}
&~& z_{0} = 0 , ~~~~(n = m = 0) 
\nonumber \\
&~& z_{1} = \frac{1}{3}(2e_1 + e_2) 
= \frac{1}{\sqrt{3}} e^{\pi i/6} ,~~~~(n = 1,  m =0) 
\nonumber \\
&~& z_{2} = \frac{1}{3}(e_1 + 2 e_2) = \frac{1}{\sqrt{3}} e^{\pi i/2} ,~~~~(n = m = 1) 
\label{z012}
\end{eqnarray}
where we take $\theta = \omega$.
The $Z_3$ transformations around the fixed points $z_0$, $z_1$ and $z_2$ and
shifts by $e_1$ and $e_2$ are defined by
\begin{eqnarray}
&~& s_0: z \to \theta z = \omega z , ~~ s_1: z - z_1 \to \theta (z - z_1) , ~~
s_2: z - z_2 \to \theta (z - z_2) , 
\nonumber \\
&~& t_1: z \to z + e_1 = z + 1, ~~ t_2: z \to z + e_2 = z + \omega .
\label{Z3&Shifts}
\end{eqnarray}
Using Eq. (\ref{z012}), the operations $s_1$ and $s_2$ are written as
\begin{eqnarray}
&~& s_1: z \to \theta z + e_1 = \omega z + 1 ,
\nonumber \\
&~& s_2: z \to \theta z + e_1 + e_2 = \omega z + 1 + \omega = \omega z - \bar{\omega} ,
\label{s12}
\end{eqnarray}
where $\bar{\omega}= e^{-2\pi i/3} = e^{4\pi i/3}$ and we use the relation $1+\omega + \bar{\omega}=0$.
Among the above operations, the following relations hold:
\begin{eqnarray}
&~& s_0^3 = s_1^3 = s_2^3 
= s_2 s_0 s_1 = s_0 s_1 s_2 = s_1 s_2 s_0 = I , 
\nonumber \\
&~& s_1 = t_1 s_0 , ~~ s_2 = t_2 t_1 s_0 , ~~ t_1 t_2 = t_2 t_1 ,
\label{Z3-rel}
\end{eqnarray}
where $I$ is the identity operation.
$s_2$, $t_1$ and $t_2$ are not independent of $s_0$ and $s_1$.

On $T^2/Z_3$, the point $z$ is identified by the points
$z+e_1$, $z+e_2$ and $\theta z$, but all six-dimensional bulk fields do not necessarily 
take identical values at these points.
Let the bulk field $\Phi(x,z,\bar{z})$ be a multiplet of some transformation group $G$ and
the Lagrangian density $\mathcal{L}$ be invariant under the transformation 
$\Phi(x,z,\bar{z}) \to \Phi'(x,z,\bar{z})=T_{\Phi}\Phi(x,z,\bar{z})$ 
such that 
\begin{eqnarray}
\mathcal{L}(\Phi(x,z,\bar{z})) = \mathcal{L}(\Phi'(x,z,\bar{z})) ,
\label{L-inv}
\end{eqnarray}
where $T_{\Phi}$ is a representation matrix of $G$ on $\Phi$.
When we require $\mathcal{L}$ to be single-valued on $M^4 \times (T^2/Z_3)$, i.e.,
\begin{eqnarray}
&~& \mathcal{L}(\Phi(x,z,\bar{z})) = \mathcal{L}(\Phi(x,z+1,\bar{z}+1)) 
= \mathcal{L}(\Phi(x,z+\omega,\bar{z}+\bar{\omega}))
\nonumber \\ 
&~& ~~~~~~~~~~~~~~~~ = \mathcal{L}(\Phi(x,\omega z,\bar{\omega}\bar{z})) ,
\label{L-single}
\end{eqnarray}
the field can be identified such that\footnote{
If fields and their superpartners yield different BCs,
the Scherk-Schwarz mechanism can work.\cite{SS}}
\begin{eqnarray}
\hspace{-0.5cm} &~& \Phi(x, \omega z, \bar{\omega} \bar{z}) = T_{\Phi}[\hat{\Theta}_0] \Phi(x, z, \bar{z}) , ~~
\Phi(x, \omega z +1, \bar{\omega} \bar{z}+1) = T_{\Phi}[\hat{\Theta}_1] \Phi(x, z, \bar{z}) ,
\nonumber \\
\hspace{-0.5cm} &~& \Phi(x, \omega z +1+\omega, \bar{\omega} \bar{z}+1+\bar{\omega}) 
= T_{\Phi}[\hat{\Theta}_2] \Phi(x, z, \bar{z}) , 
\nonumber \\
\hspace{-0.5cm} &~& \Phi(x, z +1, \bar{z}+1) = T_{\Phi}[\hat{\Xi}_1] \Phi(x, z, \bar{z}) , ~~
\Phi(x, z + \omega, \bar{z}+\bar{\omega}) = T_{\Phi}[\hat{\Xi}_2] \Phi(x, z, \bar{z}) ,
\label{BCs-Phi}
\end{eqnarray}
where $T_{\Phi}[\hat{\Theta}_0]$, $T_{\Phi}[\hat{\Theta}_1]$, $T_{\Phi}[\hat{\Theta}_2]$, 
$T_{\Phi}[\hat{\Xi}_1]$ and $T_{\Phi}[\hat{\Xi}_2]$
represent appropriate representation matrices, including an arbitrary $Z_3$ phase factor.  
The counterparts of Eq. (\ref{Z3-rel}) are given by
\begin{eqnarray}
&~& T_{\Phi}[\hat{\Theta}_0]^3 = T_{\Phi}[\hat{\Theta}_1]^3 = T_{\Phi}[\hat{\Theta}_2]^3 
= T_{\Phi}[\hat{\Theta}_2] T_{\Phi}[\hat{\Theta}_0] T_{\Phi}[\hat{\Theta}_1] 
= T_{\Phi}[\hat{\Theta}_0] T_{\Phi}[\hat{\Theta}_1] T_{\Phi}[\hat{\Theta}_2] 
\nonumber \\
&~& ~~~ = T_{\Phi}[\hat{\Theta}_1] T_{\Phi}[\hat{\Theta}_2] T_{\Phi}[\hat{\Theta}_0] = I , 
\nonumber \\
&~& T_{\Phi}[\hat{\Theta}_1] = T_{\Phi}[\hat{\Xi}_1] T_{\Phi}[\hat{\Theta}_0] , ~~
T_{\Phi}[\hat{\Theta}_2] = T_{\Phi}[\hat{\Xi}_2] T_{\Phi}[\hat{\Xi}_1] T_{\Phi}[\hat{\Theta}_0] ,
\nonumber \\
&~& T_{\Phi}[\hat{\Xi}_2] T_{\Phi}[\hat{\Xi}_1] = T_{\Phi}[\hat{\Xi}_1] T_{\Phi}[\hat{\Xi}_2] ,
\label{Phi-rel}
\end{eqnarray}
where $I$ stands for the unit matrix.
For instance, if $\Phi$ belongs to the fundamental representation of the $SU(N)$ gauge group
and a singlet under $Z_3$ transformation,
then $T_\Phi [\hat{\Theta}_0] \Phi$ is $\eta_0 \Theta_0 \Phi$, where $\Theta_0$ is
a $U(N)$ matrix, i.e., $\Theta_0^{\dagger} = \Theta_0^{2} = \Theta_0^{-1}$,
and $\eta_0$ is an intrinsic phase factor given by a qubic root.
The same property applies to $T_\Phi [\hat{\Theta}_1]$ and $T_\Phi [\hat{\Theta}_2]$.
By using Eq. (\ref{Phi-rel}), the representations of shifts are given by those of $Z_3$ rotations such that
\begin{eqnarray}
&~& T_{\Phi}[\hat{\Xi}_1] = T_{\Phi}[\hat{\Theta}_1] T_{\Phi}[\hat{\Theta}_0]^{{\dagger}} 
=  T_{\Phi}[\hat{\Theta}_1] T_{\Phi}[\hat{\Theta}_0]^{2} ,
\nonumber  \\
&~& T_{\Phi}[\hat{\Xi}_2] = T_{\Phi}[\hat{\Theta}_2] T_{\Phi}[\hat{\Theta}_1]^{{\dagger}} 
=  T_{\Phi}[\hat{\Theta}_2] T_{\Phi}[\hat{\Theta}_1]^{2} .
\label{T-rel}
\end{eqnarray}
Furthermore the representation of $s_2$ is given by other $Z_3$ rotations such that
\begin{eqnarray}
&~& T_{\Phi}[\hat{\Theta}_2] = T_{\Phi}[\hat{\Theta}_1]^{\dagger} T_{\Phi}[\hat{\Theta}_0]^{\dagger} 
=  T_{\Phi}[\hat{\Theta}_1]^2 T_{\Phi}[\hat{\Theta}_0]^{2} .
\label{T-rel2}
\end{eqnarray}
Hereafter, we use two kinds of $Z_3$ rotations, $s_0$ and $s_1$, as independent operations.

Let $G$ be a direct product of a gauge group and a $\lq$flavor' group.
The BCs imposed on the six-dimensional gauge field $A_M(x,z,\bar{z})$ are given by
\begin{eqnarray}
&~& A_\mu(x, \omega z, \bar{\omega} \bar{z}) = \Theta_0 A_\mu(x, z, \bar{z}) \Theta_0^{\dagger} , ~~
A_z(x, \omega z, \bar{\omega} \bar{z}) = \bar{\omega} \Theta_0 A_z(x, z, \bar{z}) \Theta_0^{\dagger} ,
\nonumber \\
&~& A_{\bar{z}}(x, \omega z, \bar{\omega} \bar{z}) 
= \omega \Theta_0 A_{\bar{z}}(x, z, \bar{z}) \Theta_0^{\dagger} ,
\label{A-z0} \\
&~& A_\mu(x, \omega z + 1, \bar{\omega} \bar{z} +1) = \Theta_1 A_\mu(x, z, \bar{z}) \Theta_1^{\dagger} , 
\nonumber \\
&~& A_z(x, \omega z +1 , \bar{\omega} \bar{z} +1) = \bar{\omega} \Theta_1 A_z(x, z, \bar{z}) \Theta_1^{\dagger} ,
\nonumber \\
&~& A_{\bar{z}}(x, \omega z +1 , \bar{\omega} \bar{z} + 1) 
= \omega \Theta_1 A_{\bar{z}}(x, z, \bar{z}) \Theta_1^{\dagger} ,
\label{A-z1} \\
&~& A_\mu(x, \omega z + 1 + \omega , \bar{\omega} \bar{z} +1 + \bar{\omega}) 
= \Theta_2 A_\mu(x, z, \bar{z}) \Theta_2^{\dagger} , 
\nonumber \\
&~& A_z(x, \omega z +1 +\omega , \bar{\omega} \bar{z} +1+ \bar{\omega}) 
= \bar{\omega} \Theta_2 A_z(x, z, \bar{z}) \Theta_2^{\dagger} ,
\nonumber \\
&~& A_{\bar{z}}(x, \omega z +1+ \omega , \bar{\omega} \bar{z} + 1+ \bar{\omega}) 
= \omega \Theta_2 A_{\bar{z}}(x, z, \bar{z}) \Theta_2^{\dagger} ,
\label{A-z2} \\
&~& A_M(x, z+1, \bar{z}+1) = \Xi_1 A_M(x, z, \bar{z}) \Xi_1^{\dagger} , 
\nonumber \\
&~& A_M(x, z+\omega, \bar{z}+\bar{\omega}) = \Xi_2 A_M(x, z, \bar{z}) \Xi_2^{\dagger} ,
\label{A-t}
\end{eqnarray}
where $(\Theta_0, \Theta_1, \Theta_2, \Xi_1, \Xi_2)$ are representation matrices of the gauge group (times $U(1)$s).
These BCs are consistent with the gauge covariance of the derivative $D_M = \partial_M + ig A_M(x,z,\bar{z})$,
where $g$ is a gauge coupling constant.
For the bulk scalar field $\phi(x,z,\bar{z})$, which is a singlet under $Z_3$ transformation, BCs are given by
\begin{eqnarray}
&~& \phi(x, \omega z, \bar{\omega} \bar{z})  = T_{\phi}[\Theta_0] \phi(x, z, \bar{z}) , 
\nonumber \\
&~& \phi(x, \omega z + 1, \bar{\omega} \bar{z} + 1)  
 = T_{\phi}[\Theta_1] \phi(x, z, \bar{z}) , ~~
\nonumber \\
&~& \phi(x, \omega z + 1 + \omega, \bar{\omega} \bar{z} + 1 + \bar{\omega})  
 = T_{\phi}[\Theta_2] \phi(x, z, \bar{z}) ,
\nonumber \\
&~& \phi(x, z + 1, \bar{z} + 1) = T_{\phi}[\Xi_1] \phi(x, z, \bar{z}) , 
\nonumber \\
&~& \phi(x, z + \omega, \bar{z} + \bar{\omega}) = T_{\phi}[\Xi_2] \phi(x, z, \bar{z}) .
\label{phi-BC}
\end{eqnarray}
For a set of scalar fields $\phi^A(x,z,\bar{z})$ $(A=1, 2, 3)$ that form a triplet under $Z_3$ transformation, 
these BCs are given by
\begin{eqnarray}
&~& \phi^A(x, \omega z, \bar{\omega} \bar{z})  = T_{\phi^A}[\hat{\Theta}_0] \phi^A(x, z, \bar{z}) 
= \sum_{B=1}^3 {(\cal{X})^A}_B T_{\phi^B}[\Theta_0] \phi^B(x, z, \bar{z}) , 
\nonumber \\
&~& \phi^A(x, \omega z + 1, \bar{\omega} \bar{z} + 1)  
\nonumber \\
&~& ~~~~~~~~~~~~~~~~~ = T_{\phi^A}[\hat{\Theta}_1] \phi^A(x, z, \bar{z}) 
 = \sum_{B=1}^3 {\big( e^{-2\pi i \gamma \cal{Y}} \cal{X} \big)^A}_B T_{\phi^B}[\Theta_1] \phi^B(x, z, \bar{z}) , ~~
\nonumber \\
&~& \phi^A(x, \omega z + 1 + \omega, \bar{\omega} \bar{z} + 1 + \bar{\omega}) 
\nonumber \\
&~& ~~~~~~~~~~~~~~~~~ = T_{\phi^A}[\hat{\Theta}_2] \phi^A(x, z, \bar{z}) 
= \sum_{B=1}^3 {\big( e^{2\pi i \gamma \cal{Y}_{\omega}} \cal{X} \big)^A}_B 
T_{\phi^B}[\Theta_2] \phi^B(x, z, \bar{z}) ,
\nonumber \\
&~& \phi^A(x, z + 1, \bar{z} + 1) 
\nonumber \\
&~& ~~~~~~~~~~~~~~~~~  = T_{\phi^A}[\hat{\Xi}_1] \phi^A(x, z, \bar{z}) 
 = \sum_{B=1}^3 {\big( e^{-2\pi i \gamma \cal{Y}} \big)^A}_B T_{\phi^B}[\Xi_1] \phi^B(x, z, \bar{z}) , 
\nonumber \\
&~& \phi^A(x, z + \omega, \bar{z} + \bar{\omega}) 
\nonumber \\
&~& ~~~~~~~~~~~~~~~~~ = T_{\phi^A}[\hat{\Xi}_2] \phi^A(x, z, \bar{z})  
= \sum_{B=1}^3 {\big( e^{-2\pi i \gamma \cal{Y}_{\bar{\omega}}} \big)^A}_B 
T_{\phi^B}[\Xi_2] \phi^B(x, z, \bar{z}) ,
\label{phiA-BC}
\end{eqnarray}
where $\cal{X}$, $\cal{Y}$, $\cal{Y}_{\omega}$ and $\cal{Y}_{\bar{\omega}}$ 
are $3 \times 3$ matrices 
and the parameter $\gamma$ can take an arbitrary real value.
Here, the cyclic group $Z_3$ is a discrete subgroup of the $\lq$flavor' group.
For the $Z_3$ singlet Dirac field $\psi(x,z,\bar{z})$ defined in the bulk, 
the gauge invariance of the kinetic energy term requires the following BCs:
\begin{eqnarray}
&~& \psi(x, \omega z, \bar{\omega} \bar{z})  = T_{\psi}[\Theta_0] S_0 \psi(x, z, \bar{z}) , 
\nonumber \\ 
&~& \psi(x, \omega z + 1, \bar{\omega} \bar{z} + 1)  =  
T_{\psi}[\Theta_1] S_1 \psi(x, z, \bar{z}) ,
\nonumber \\
&~& \psi(x, \omega z + 1 + \omega, \bar{\omega} \bar{z} + 1 + \bar{\omega})  
= T_{\psi}[\Theta_2] S_2 \psi(x, z, \bar{z}) ,
\nonumber \\ 
&~& \psi(x, z + 1, \bar{z} + 1)  = T_{\psi}[\Xi_1] S_1 S_0^2 \psi(x, z, \bar{z}) ,
\nonumber \\
&~& \psi(x, z + \omega, \bar{z} + \bar{\omega})  
=  T_{\psi}[\Xi_2] S_2 S_1^2 \psi(x, z, \bar{z}) ,
\label{psi-BC}
\end{eqnarray}
where $S_i$ $(i = 0, 1, 2)$ are $8 \times 8$ matrices acting on the Dirac spinor given by
\begin{eqnarray}
S_i = I_{4 \times 4} \otimes \frac{1}{2}
\left(
\begin{array}{cc}
1 & 0 \\
0 & \omega 
\end{array}
\right) 
= \frac{1}{8}\left(\Gamma^z \Gamma^{\bar{z}} + \omega \Gamma^{\bar{z}} \Gamma^z\right) .
\label{Si}
\end{eqnarray}
Here, $I_{4 \times 4}$ is the $4 \times 4$ unit matrix, and 
arbitrary $Z_3$ phase factors are absorbed by the intrinsic ones $\eta_i$.
We use the following representation for six-dimensional gamma matrices:
\begin{eqnarray}
&~& \Gamma^{\mu} = \gamma^{\mu} \otimes \sigma_3 , ~~
\Gamma^{5} = I_{4 \times 4}  \otimes \sigma_1 , ~~ 
\Gamma^{6} = I_{4 \times 4}  \otimes \sigma_2 ,
\nonumber \\ 
&~& \Gamma^{z} \equiv \Gamma^{5} + i \Gamma^{6} = 2 I_{4 \times 4}  \otimes \sigma_{+} , ~~
\Gamma^{\bar{z}} \equiv \Gamma^{5} - i \Gamma^{6} = 2 I_{4 \times 4}  \otimes \sigma_{-}  .
\label{Gamma}
\end{eqnarray}
The following relations hold:
\begin{eqnarray}
&~& \Gamma^{\mu} S_i = S_i \Gamma^{\mu} , ~~
\Gamma^{z} S_i = \omega S_i \Gamma^{z} , ~~
\Gamma^{\bar{z}} S_i = \bar{\omega} S_i \Gamma^{\bar{z}} .
\label{GS-rel}
\end{eqnarray}
The BCs for a $Z_3$ triplet Dirac field are similarly given.

In this way, we find that BCs in gauge theories on $T^2/Z_3$ are specified by
$(\Theta_0, \Theta_1,$ $\gamma)$ and additional $Z_3$ phase factors.

\subsection{Residual gauge invariance and equivalence classes}

Given the BCs $(\Theta_0, \Theta_1, \Theta_2, \Xi_1, \Xi_2, \gamma)$, 
there still remains residual gauge invariance.  
Under gauge transformation with the transformation function $\Omega = \Omega(x, z, \bar{z})$, 
fields are transformed as
\begin{eqnarray}
&~& A_M \to {A'}_M = \Omega A_M  \Omega^{\dagger} - {i \over g}\Omega \partial_M \Omega^{\dagger} , ~~
\phi \to {\phi'} = T_{\phi} [\Omega] \phi , 
\nonumber \\
&~& \phi^A \to {\phi'}^A = T_{\phi^A} [\Omega] \phi^A , ~~
\psi^A \to {\psi'}^A = T_{\psi^A}[\Omega] \psi^A , 
\label{gauge-transf}
\end{eqnarray}
where $A'_M(x,z,\bar{z})$ satisfies, instead of Eqs. (\ref{A-z0}) -- (\ref{A-t}),
\begin{eqnarray}
&~& A'_{\mu}(x, \omega z, \bar{\omega} \bar{z}) = \Theta'_0 A'_{\mu}(x, z, \bar{z}) {\Theta'}_0^{\dagger} 
 - \frac{i}{g} \Theta'_0 \partial_\mu {\Theta'}_0^{\dagger} , 
\nonumber \\
&~& A'_z(x, \omega z, \bar{\omega} \bar{z}) = \bar{\omega} \left(\Theta'_0 A'_z(x, z, \bar{z}) {\Theta'}_0^{\dagger}
 - \frac{i}{g} \Theta'_0 \partial_z {\Theta'}_0^{\dagger}\right) ,
\nonumber \\
&~& A'_{\bar{z}}(x, \omega z, \bar{\omega} \bar{z}) = \omega \left(\Theta'_0 A'_{\bar{z}}(x, z, \bar{z}) {\Theta'}_0^{\dagger} 
 - \frac{i}{g} \Theta'_0 \partial_{\bar{z}} {\Theta'}_0^{\dagger}\right) ,
\label{A'-z0} \\
&~& A'_{\mu}(x, \omega z + 1, \bar{\omega} \bar{z} +1) = \Theta'_1 A_{\mu}(x, z, \bar{z}) {\Theta'}_1^{\dagger} 
 - \frac{i}{g} \Theta'_1 \partial_\mu {\Theta'}_1^{\dagger} , 
\nonumber \\
&~& A'_z(x, \omega z +1 , \bar{\omega} \bar{z} +1) = \bar{\omega} \left(\Theta'_1 A'_z(x, z, \bar{z}) {\Theta'}_1^{\dagger} 
 - \frac{i}{g} \Theta'_1 \partial_z {\Theta'}_1^{\dagger}\right) ,
\nonumber \\
&~& A'_{\bar{z}}(x, \omega z +1 , \bar{\omega} \bar{z} + 1) 
= \omega \left(\Theta'_1 A'_{\bar{z}}(x, z, \bar{z}) {\Theta'}_1^{\dagger} 
 - \frac{i}{g} \Theta'_1 \partial_{\bar{z}} {\Theta'}_1^{\dagger}\right) ,
\label{A'-z1} \\
&~& A'_{\mu}(x, \omega z + 1 + \omega , \bar{\omega} \bar{z} +1 + \bar{\omega}) 
= \Theta'_2 A'_{\mu}(x, z, \bar{z}) {\Theta'}_2^{\dagger}  - \frac{i}{g} \Theta'_2 \partial_\mu {\Theta'}_2^{\dagger}, 
\nonumber \\
&~& A'_z(x, \omega z +1 +\omega , \bar{\omega} \bar{z} +1+ \bar{\omega}) 
= \bar{\omega} \left(\Theta'_2 A'_z(x, z, \bar{z}) {\Theta'}_2^{\dagger} 
 - \frac{i}{g} \Theta'_2 \partial_z {\Theta'}_2^{\dagger}\right) ,
\nonumber \\
&~& A'_{\bar{z}}(x, \omega z +1+ \omega , \bar{\omega} \bar{z} + 1+ \bar{\omega}) 
= \omega \left(\Theta'_2 A'_{\bar{z}}(x, z, \bar{z}) {\Theta'}_2^{\dagger} 
 - \frac{i}{g} \Theta'_2 \partial_{\bar{z}} {\Theta'}_2^{\dagger}\right) ,
\label{A'-z2} \\
&~& A'_M(x, z+1, \bar{z}+1) = {\Xi}'_1 A'_M(x, z, \bar{z}) {\Xi'}_1^{\dagger} 
 - \frac{i}{g} \Xi'_1 \partial_M {\Xi'}_1^{\dagger} , 
\nonumber \\
&~& A'_M(x, z+\omega, \bar{z}+\bar{\omega}) = \Xi'_2 A_M(x, z, \bar{z}) {\Xi'}_2^{\dagger} 
 - \frac{i}{g} \Xi'_2 \partial_M {\Xi'}_2^{\dagger} .
\label{A'-t}
\end{eqnarray}
Here, $\Theta'_0$, $\Theta'_1$, $\Theta'_2$, ${\Xi'}_1$ and ${\Xi'}_2$ are given by
\begin{eqnarray}
&~& \Theta'_0 = \Omega(x, \omega z, \bar{\omega} \bar{z}) \Theta_0 \Omega^{\dagger} (x, z, \bar{z}) , 
\nonumber \\
&~& \Theta'_1 = \Omega(x, \omega z + 1, \bar{\omega} \bar{z} + 1) \Theta_1 \Omega^{\dagger} (x, z, \bar{z}) ,
\nonumber \\
&~& \Theta'_2 = \Omega(x, \omega z + 1 + \omega, \bar{\omega} \bar{z} + 1 + \bar{\omega}) \Theta_2 
\Omega^{\dagger} (x, z, \bar{z}) ,
\nonumber \\
&~& {\Xi'}_1 = \Omega(x,  z + 1, \bar{z} + 1 ) \Xi_1 \Omega^{\dagger} (x, z, \bar{z}) , 
\nonumber \\
&~& {\Xi'}_2 = \Omega(x,  z + \omega, \bar{z} + \bar{\omega}) \Xi_2 \Omega^{\dagger} (x, z, \bar{z}) .
\label{BC'}
\end{eqnarray}
The scalar fields ${\phi'}(x, z, \bar{z})$ and ${\phi'}^A(x,z,\bar{z})$ and the Dirac fermion ${\psi'}(x,z,\bar{z})$ 
satisfy relations similar to Eqs. (\ref{phi-BC}) -- (\ref{psi-BC}), 
where $(\Theta_0, \Theta_1, \Theta_2, \Xi_1, \Xi_2, \gamma)$ is replaced by 
$(\Theta'_0, \Theta'_1, \Theta'_2, {\Xi'}_1, {\Xi'}_2, \gamma)$.

The residual gauge invariance of the BCs is given by gauge
transformations that preserve the given BCs, $\Theta'_0 = \Theta_0$, $\Theta'_1 = \Theta_1$, 
$\Theta'_2 = \Theta_2$, $\Xi'_1 = \Xi_1$ and $\Xi'_2 = \Xi_2$:
\begin{eqnarray}
&~& \Omega(x, \omega z, \bar{\omega} \bar{z}) \Theta_0  = \Theta_0 \Omega (x, z, \bar{z}) , 
\nonumber \\
&~& \Omega(x, \omega z +1 , \bar{\omega} \bar{z} +1) \Theta_1  = \Theta_1 \Omega (x, z, \bar{z}) , 
\nonumber \\
&~& \Omega(x, \omega z + 1 + \omega, \bar{\omega} \bar{z} + 1 \bar{\omega}) \Theta_2  = \Theta_2 \Omega (x, z, \bar{z}) , 
\label{res-gauge-theta} \\
&~& \Omega(x, z +1 , \bar{z} +1) \Xi_1  = \Xi_1 \Omega (x, z, \bar{z}) , 
\nonumber \\ 
&~& \Omega(x, z +\omega , \bar{z} + \bar{\omega}) \Xi_2  = \Xi_2 \Omega (x, z, \bar{z}) .
\label{res-gauge-t}
\end{eqnarray}
We refer to the residual gauge invariance of BCs as the gauge symmetry of BCs. The low-energy gauge symmetry of BCs 
is derived from the following relations that are independent of extradimensional
coordinates:
\begin{eqnarray}
&~& \Omega(x) \Theta_0  = \Theta_0 \Omega (x) , ~~ \Omega(x) \Theta_1 = \Theta_1 \Omega (x) , ~~
\Omega(x) \Theta_2  =  \Theta_2 \Omega (x) ,
\label{LES-theta}\\
&~& \Omega(x) \Xi_1  = \Xi_1 \Omega (x) , ~~ \Omega(x) \Xi_2 = \Xi_2 \Omega (x) .
\label{LES-t}
\end{eqnarray}
The symmetry is generated by generators that commute with $\Theta_0$ and $\Theta_1$.

Theories with different BCs should be equivalent in terms of physics content
if they are connected by gauge transformations. 
The key observation is that physics should not depend on the gauge chosen.
If $(\Theta'_0, \Theta'_1, \Theta'_2, \Xi'_1, \Xi'_2)$ satisfies the conditions
\begin{eqnarray}
&~& \partial_M \Theta'_0 = 0 , ~~ \partial_M \Theta'_1 = 0 , ~~  \partial_M \Theta'_2 = 0 , ~~
\partial_M \Xi'_1 = 0 , ~~ \partial_M \Xi'_2 = 0 ,
\label{equiv1}\\
&~& {\Theta'}_0^3 = {\Theta'}_1^3 = {\Theta'}_2^3 
= \Theta'_2 \Theta'_0 \Theta'_1 = \Theta'_0 \Theta'_1 \Theta'_2 = \Theta'_1 \Theta'_2 \Theta'_0 = I ,
\label{equiv2}
\end{eqnarray}
then the two sets of BCs are equivalent:
\begin{eqnarray}
(\Theta'_0, \Theta'_1, \Theta'_2, \Xi'_1, \Xi'_2) \sim (\Theta_0, \Theta_1, \Theta_2, \Xi_1, \Xi_2) .
\label{equ}
\end{eqnarray}
The equivalence relation (\ref{equ}) defines equivalence classes of BCs.
Here, we illustrate the change of BCs under a singular gauge transformation.
Let us consider an $SU(3)$ gauge theory with 
$(\Theta_0, \Theta_1, \Theta_2, \Xi_1, \Xi_2) = (X, X, X, I, I)$.
Here, $X$ and $I$ are given by
\begin{eqnarray}
X = 
\left(
\begin{array}{ccc}
1 & 0 & 0 \\
0 & \omega & 0 \\
0 & 0 & \omega^2 
\end{array}
\right), ~~
I = 
\left(
\begin{array}{ccc}
1 & 0 & 0 \\
0 & 1 & 0 \\
0 & 0 & 1 
\end{array}
\right) .
\label{X}
\end{eqnarray}
We carry out the gauge transformation defined by
\begin{eqnarray}
\Omega = \exp\left(ia\left(Y_{+}^1 z + Y_{-}^1 \bar{z}\right)\right) ,
\label{Omega}
\end{eqnarray}
where $a$ is a real number, and $Y_{+}^1$ and $Y_{-}^1$ are defined by
\begin{eqnarray}
Y_{+}^1 = 
\left(
\begin{array}{ccc}
0 & 1 & 0 \\
0 & 0 & 1 \\
1 & 0 & 0 
\end{array}
\right), ~~
Y_{-}^1 = 
\left(
\begin{array}{ccc}
0 & 0 & 1 \\
1 & 0 & 0 \\
0 & 1 & 0 
\end{array}
\right) .
\label{Y}
\end{eqnarray}
Then we find the equivalence relation
\begin{eqnarray}
(X, X, X, I, I) \sim (X, e^{iaY}X, e^{-iaY_{\omega}}X, e^{iaY}I, e^{iaY_{\bar{\omega}}}I) ,
\label{equX}
\end{eqnarray}
where $Y$, $Y_{\omega}$ and $Y_{\bar{\omega}}$ are defined by
\begin{eqnarray}
\hspace{-0.7cm} Y = 
\left(
\begin{array}{ccc}
0 & 1 & 1 \\
1 & 0 & 1 \\
1 & 1 & 0 
\end{array}
\right), ~~
Y_{\omega} = 
\left(
\begin{array}{ccc}
0 & \omega & \omega^2 \\
\omega^2 & 0 & \omega \\
\omega & \omega^2 & 0 
\end{array}
\right), ~~
Y_{\bar{\omega}} = 
\left(
\begin{array}{ccc}
0 & \omega^2 & \omega \\
\omega & 0 & \omega^2 \\
\omega^2 & \omega & 0 
\end{array}
\right) .
\label{Ys}
\end{eqnarray}
In particular, we have the equivalence relation
\begin{eqnarray}
(X, X, X, I, I) \sim (X, X_{\omega}, X_{\bar{\omega}}, \omega I, \omega I) ,
\label{equX1}
\end{eqnarray}
for $a = 4\pi/3$, and the equivalence relation
\begin{eqnarray}
(X, X, X, I, I) \sim (X, X_{\bar{\omega}}, X_{\omega}, \bar{\omega} I, \bar{\omega} I) ,
\label{equX2}
\end{eqnarray}
for $a = 2\pi/3$.
Here, $X_{\omega}$ and $X_{\bar{\omega}}$ are defined by
\begin{eqnarray}
X_{\omega} = 
\left(
\begin{array}{ccc}
\omega & 0 & 0 \\
0 & \omega^2 & 0 \\
0 & 0 & 1 
\end{array}
\right) , ~~
 X_{\bar{\omega}} =  
\left(
\begin{array}{ccc}
\omega^2 & 0 & 0 \\
0 & 1 & 0 \\
0 & 0 & \omega 
\end{array}
\right) .
\label{Xomega}
\end{eqnarray}
In this way, BCs can change under gauge transformations. 

The symmetry of BCs in one theory differs from that in the other, but 
two theories should describe the same physics and be equivalent
if they are related to by gauge transformations.
This equivalence is guaranteed in the Hosotani mechanism, as will be explained in the next subsection.

\subsection{Hosotani mechanism and physical symmetry}

The Hosotani mechanism\cite{H} in gauge theories defined on $T^2/Z_3$ is summarized as follows.\\
(i)  Wilson line phases are phase factors in $W_j \Xi_j$ $(j = 1, 2)$ defined by
\begin{eqnarray}
&~& W_1 \Xi_1 \equiv P \exp \Big\{ig  \int_{C_1} (A_z dz + A_{\bar{z}} d\bar{z})\Big\} \Xi_1 ,
\label{W1}\\
&~& W_2 \Xi_2 \equiv P \exp \Big\{ig  \int_{C_2} (A_z dz + A_{\bar{z}} d\bar{z})\Big\} \Xi_2 ,
\label{W2}
\end{eqnarray}
where $C_j$ are noncontractible loops on $T^2$.
The eigenvalues of $W_j\Xi_j$ are gauge-invariant and become physical degrees of freedom.
Hence, Wilson line phases cannot be gauged away and parametrize degenerate vacua at the tree level.\\
(ii) The degeneracy is, in general, lifted by quantum effects.
The physical vacuum is given by the configuration of Wilson line phases that minimizes 
the effective potential $V_{\rm eff}$.\\
(iii) If the configuration of the Wilson line phases is nontrivial,
the gauge symmetry is spontaneously broken or restored by radiative corrections.
Nonvanishing expectation values of the Wilson line phases give masses to gauge fields related to broken symmetries.
Extradimensional components of gauge fields and some matter fields also acquire masses.\\
(iv)  Two physical systems are equivalent 
if they are connected by a gauge transformation, which is a symmetry of the Lagrangian
\begin{eqnarray}
\Bigl.\mathcal{L} 
\left(\Phi(x,z,\bar{z})\right)\Bigr|_{\left(\langle A_z \rangle, \langle A_{\bar{z}} \rangle, \Theta_0, \Theta_1\right)} 
= \Bigl.\mathcal{L}
(\Phi'(x,z,\bar{z}))\Bigr|_{\left(\langle A'_z \rangle, \langle A'_{\bar{z}} \rangle, \Theta'_0, \Theta'_1\right)} 
\label{L-gaugeinv}
\end{eqnarray}
and is also preserved in the effective potential
\begin{eqnarray}
V_\text{eff} \left(\langle A_z \rangle, \langle A_{\bar{z}} \rangle, \Theta_0, \Theta_1 \right) 
= V_\text{eff} \left(\langle A'_z \rangle, \langle A'_{\bar{z}} \rangle, \Theta'_0, \Theta'_1\right).
\end{eqnarray}
The physical symmetries, parameters and spectrum are determined 
by the combination of BCs and the expectation value of Wilson line phases.\footnote{
The dynamical rearrangement of QCD theta parameter was studied in a five-dimensional gauge theory
with a mixed Chern-Simons term.\cite{HKO}}

Let us explain the last part of the mechanism in detail 
and how physical symmetry is determined.
Dynamical phases are associated with the zero modes ($z$-independent modes) of $A_z$ and $A_{\bar{z}}$ given by
\begin{eqnarray}
\left\{ \sum_p A_z^p T^p + \sum_{\bar{p}} A_{\bar{z}}^{\bar{p}} T^{\bar{p}} ; ~~ T^p, T^{\bar{p}} \in {\cal H}_W \right\} ,
\label{DWL}
\end{eqnarray} 
where ${\cal H}_W$ is a set of generators that satisfy:
\begin{eqnarray}
{\cal H}_W = \left\{T^p, T^{\bar{p}} ~;~ T^p \Theta_i = \bar{\omega} \Theta_i T^p , ~~ 
T^{\bar{p}} \Theta_i = \omega \Theta_i T^{\bar{p}} ,~~ i=0,1,2 \right\} .
\label{HW}
\end{eqnarray}
The potential for $A_z(x)$ and $A_{\bar{z}}(x)$ at the tree level is given by
\begin{eqnarray}
V_{\rm tree} = \frac{1}{2} \mbox{tr}[D_z, D_{\bar{z}}]^2  = \frac{g^2}{2} \mbox{tr}[A_z, A_{\bar{z}}]^2  .
\label{Vtree}
\end{eqnarray}
$V_{\rm tree}$ takes a minimum when the expectation value of field strength $F_{z\bar{z}}$ vanishes.
Suppose that, for $(\Theta_0, \Theta_1, \Theta_2, \Xi_1, \Xi_2, \gamma)$, 
$V_{\rm eff}$ is minimized at $\langle  A_z \rangle$ and $\langle  A_{\bar{z}} \rangle$ such that 
$\langle F_{z\bar{z}} \rangle = 0$ and $W_1 \neq I$ and/or $W_2 \neq I$.
Perform the gauge transformation given by
$\Omega = \exp\{i g (\langle A_z \rangle z + \langle A_{\bar{z}} \rangle \bar{z})\}$.
This transforms $\langle A_z \rangle$ and $\langle A_{\bar{z}} \rangle$ 
into $\langle {A'}_z \rangle = \langle A'_{\bar{z}} \rangle = 0$.
With this transformation, BCs change to
\begin{eqnarray}
&~& (\Theta'_0, \Theta'_1, \Theta'_2, \Xi'_1, \Xi'_2, \gamma)
 = (\Theta_0, \Omega(e_1) \Theta_1, \Omega(e_1+e_2) \Theta_2, \Omega(e_1) \Xi_1, \Omega(e_2) \Xi_2, \gamma) 
\nonumber \\
&~& ~~~~~ \equiv (\Theta_0^{\rm sym}, \Theta_1^{\rm sym}, \Theta_2^{\rm sym}, \Xi_1^{\rm sym}, \Xi_2^{\rm sym}, \gamma) ,
\label{Theta-sym}
\end{eqnarray}
where $\Omega(e_1)$, $\Omega(e_2)$ and $\Omega(e_1+e_2)$ are defined by
\begin{eqnarray}
&~& \Omega(e_1) = \exp\{i g (\langle A_z \rangle + \langle A_{\bar{z}} \rangle)\} , ~~
\Omega(e_2) = \exp\{i g (\omega \langle A_z \rangle + \bar{\omega} \langle A_{\bar{z}} \rangle)\} ,
\nonumber \\
&~& \Omega(e_1 + e_2) = \exp\{- i g (\bar{\omega} \langle A_z \rangle + \omega \langle A_{\bar{z}} \rangle)\} .
\label{Omega-ei}
\end{eqnarray}
Because the expectation values of $A'_z$ and $A'_{\bar{z}}$ vanish in the new gauge, 
the physical symmetry is spanned by the generators
that commute with $(\Theta_0^{\rm sym}, \Theta_1^{\rm sym})$: 
\begin{eqnarray}
{\cal H}^{\rm sym} =  \left\{ T^{\alpha} ~;~ [T^{\alpha}, \Theta_0^{\rm sym}] 
= [T^{\alpha}, \Theta_1^{\rm sym}] = 0 \right\} .
\label{Hsym}
\end{eqnarray}
The group generated by ${\cal H}^{\rm sym}$ defines the unbroken physical symmetry of the theory.

\section{Mode expansions and effective potential}

\subsection{Mode expansions of six-dimensional fields}

Fields are classified  as either $Z_3$ singlets or $Z_3$ triplets on $T^2/Z_3$.
There are nine kinds of $Z_3$ singlet fields 
denoted by $\displaystyle{\phi^{(\theta_0 \theta_1 \theta_2)}(x, z, \bar{z})}$
where $\theta_i$ are eigenvalues of $\Theta_i$.\footnote{
For convenience, $\theta_2$ is denoted
though it is not an independent parameter.
Note that the relation $\theta_0 \theta_1 \theta_2 = 1$
stems from $T_{\Phi}[\Theta_0] T_{\Phi}[\Theta_1] T_{\Phi}[\Theta_2]= I$.}
The mode expansions of $\displaystyle{\phi^{(\theta_0 \theta_1 \theta_2)}(x, z, \bar{z})}$ are given by
\begin{eqnarray}
&~& \phi^{(111)} (x, z, \bar{z}) = \phi_{0,0}(x)  + {\sum_{n, m}}' \phi_{n, m}(x) f_{n, m}^{(0)}(z, \bar{z}) ,
\nonumber \\
&~& \phi^{(1 \omega \bar{\omega})} (x, z, \bar{z}) 
= \sum_{n, m} \phi_{n, m}(x) f_{n+\frac{1}{3}, m+\frac{1}{3}}^{(0)}(z, \bar{z}) ,
\nonumber \\
&~& \phi^{(1 \bar{\omega} \omega)} (x, z, \bar{z}) 
= \sum_{n, m} \phi_{n, m}(x) f_{n+\frac{2}{3}, m+\frac{2}{3}}^{(0)}(z, \bar{z}) ,
\label{exp1} \\
&~& \phi^{(\omega \omega \omega)} (x, z, \bar{z}) 
= {\sum_{n, m}}' \phi_{n, m}(x) f_{n,m}^{(1)}(z, \bar{z}) ,
\nonumber \\
&~& \phi^{(\omega \bar{\omega} 1)} (x, z, \bar{z}) 
= \sum_{n, m} \phi_{n, m}(x) f_{n+\frac{1}{3}, m+\frac{1}{3}}^{(1)}(z, \bar{z}) ,
\nonumber \\
&~& \phi^{(\omega 1 \bar{\omega})} (x, z, \bar{z}) 
= \sum_{n, m} \phi_{n, m}(x) f_{n+\frac{2}{3}, m+\frac{2}{3}}^{(1)}(z, \bar{z}) ,
\label{exp2} \\
&~& \phi^{(\bar{\omega} \bar{\omega} \bar{\omega})} (x, z, \bar{z}) 
= {\sum_{n, m}}' \phi_{n, m}(x) f_{n,m}^{(2)}(z, \bar{z}) ,
\nonumber \\
&~& \phi^{(\bar{\omega} 1 \omega)} (x, z, \bar{z}) 
= \sum_{n, m} \phi_{n, m}(x) f_{n+\frac{1}{3}, m+\frac{1}{3}}^{(2)}(z, \bar{z}) ,
\nonumber \\
&~& \phi^{(\bar{\omega} \omega 1)} (x, z, \bar{z}) 
= \sum_{n, m} \phi_{n, m}(x) f_{n+\frac{2}{3}, m+\frac{2}{3}}^{(2)}(z, \bar{z}) ,
\label{exp3}
\end{eqnarray}
where $\sum'_{n, m}$ means the summation over integers $(n, m)$ excluding $n=m=0$ 
and normalization factors are absorbed by the four-dimensional fields $\phi_{n,m}(x)$.
Note that only $\phi^{(111)} (x, z, \bar{z})$ has a zero mode.
Here, $f_{n+\alpha, m+\beta}^{(i)}(z, \bar{z})$ are defined by
\begin{eqnarray}
&~& f_{n+\alpha, m+\beta}^{(0)}(z, \bar{z}) \equiv f_{n+\alpha, m+\beta}(z, \bar{z}) 
   +  f_{n+\alpha, m+\beta}(\omega z, \bar{\omega}\bar{z}) 
\nonumber \\
&~& ~~~~~~~~~~~~~~~~~~~~~~~ + f_{n+\alpha, m+\beta}(\bar{\omega} z, \omega \bar{z}) ,
\label{f0} \\
&~& f_{n+\alpha, m+\beta}^{(1)}(z, \bar{z}) \equiv \bar{\omega} f_{n+\alpha, m+\beta}(z, \bar{z}) 
   +  \omega f_{n+\alpha, m+\beta}(\omega z, \bar{\omega}\bar{z}) 
\nonumber \\
&~& ~~~~~~~~~~~~~~~~~~~~~~~ + f_{n+\alpha, m+\beta}(\bar{\omega} z, \omega \bar{z}) ,
\label{f1} \\
&~& f_{n+\alpha, m+\beta}^{(2)}(z, \bar{z}) \equiv \omega f_{n+\alpha, m+\beta}(z, \bar{z}) 
   +  \bar{\omega} f_{n+\alpha, m+\beta}(\omega z, \bar{\omega} \bar{z}) 
\nonumber \\
&~& ~~~~~~~~~~~~~~~~~~~~~~~ + f_{n+\alpha, m+\beta}(\bar{\omega} z, \omega \bar{z}) , 
\label{f2}
\end{eqnarray}
where $f_{n+\alpha,m+\beta}(z, \bar{z})$ is defined by
\begin{eqnarray}
&~& f_{n+\alpha,m+\beta}(z, \bar{z})  
\equiv \exp \left[\pi i\left\{\left(n + \alpha - \frac{n + \alpha + 2(m + \beta)}{\sqrt{3}}i \right) z\right.\right.
\nonumber \\
&~& ~~~~~~~~~~~~~~~~~~~~~~~~~~~~~~~ \left.\left. + \left(n + \alpha + 
\frac{n + \alpha + 2(m + \beta)}{\sqrt{3}}i \right) \bar{z} \right\}\right] . 
\label{f}
\end{eqnarray}
In the case of vanishing Wilson line phases, the mass squared of $\phi_{n,m}(x)$ is derived 
from the kinetic terms after compactification such that
\begin{eqnarray}
&~& M^2_{n,m}(\alpha, \beta) = \pi^2 \left((n+\alpha)^2 + \frac{1}{3}(n+\alpha+2(m+\beta))^2\right) 
\nonumber \\
&~& ~~~~~~~~~~~~~~~ = \frac{4}{3} \pi^2 \left[(n+\alpha)^2 + (n+\alpha)(m+\beta) + (m+\beta)^2\right] 
\nonumber \\
&~& ~~~~~~~~~~~~~~~ = \frac{1}{3} \left[\left(\frac{n+\alpha}{R}\right)^2 
+ \left(\frac{n+\alpha}{R}\right)\left(\frac{m+\beta}{R}\right) + \left(\frac{m+\beta}{R}\right)^2\right],
\label{mass}
\end{eqnarray}
where $\alpha$, $\beta = 0, 1/3, 2/3$ and the physical size $|e_1| = |e_2| = 2 \pi R$ is used in the final expression.

In gauge theories on $T^2/Z_3$, there is another important representation, a $Z_3$ triplet.  
The $Z_3$ triplet field $\phi^A(x,z,\bar{z})$ $(A=1,2,3)$ satisfies BCs such that
\begin{eqnarray}
&~& \phi^A(x, \omega z, \bar{\omega}\bar{z}) = \sum_{B=1}^3 {(\cal{X})^A}_B \phi^B(x, z, \bar{z}) , 
\nonumber \\
&~& \phi^A(x, \omega z + 1, \bar{\omega} \bar{z}+1) = \sum_{B=1}^3 
{\left(e^{-2\pi i \gamma \cal{Y}} \cal{X}\right)^A}_B \phi^B(x, z, \bar{z}) , 
\nonumber \\
&~& \phi^A(x, \omega z + 1+ \omega, \bar{\omega}\bar{z}+1+\bar{\omega}) = \sum_{B=1}^3 
{\left(e^{2\pi i \gamma \cal{Y}_{\omega}} \cal{X}\right)^A}_B \phi^B(x, z, \bar{z}) ,
\nonumber \\
&~& \phi^A(x, z + 1, \bar{z}+1) = \sum_{B=1}^3 
{\left(e^{-2\pi i \gamma \cal{Y}}\right)^A}_B \phi^B(x, z, \bar{z}) , 
\nonumber \\
&~& \phi^A(x, z + \omega, \bar{z}+\bar{\omega}) = \sum_{B=1}^3 
{\left(e^{-2\pi i \gamma \cal{Y}_{\bar{\omega}}}\right)^A}_B \phi^B(x, z, \bar{z}) ,
\label{BCtriplet}
\end{eqnarray}
where we take $T_{\phi^A}[\Theta_i] = I$ and $\gamma$ is a real number. 
When we take $({\cal{X}}, {\cal{Y}}, {\cal{Y}}_{\omega}, {\cal{Y}}_{\bar{\omega}}) = (X, Y, Y_{\omega}, Y_{\bar{\omega}})$,
the mode expansion of $\phi^A(x,z,\bar{z})$ is given by
\begin{eqnarray}
\phi^A(x, z, \bar{z}) = \sum_{n, m} \phi_{n, m}(x) 
\left(
\begin{array}{c}
f_{n+\gamma, m+\gamma}^{(0)}(z, \bar{z})  \\
f_{n+\gamma, m+\gamma}^{(1)}(z, \bar{z})  \\
f_{n+\gamma, m+\gamma}^{(2)}(z, \bar{z}) 
\end{array}
\right) ,
\label{exp-triplet}
\end{eqnarray}
where $\gamma$ can take an arbitrary value.
In the case of vanishing Wilson line phases, the mass squared of $\phi_{n, m}(x)$ is given by 
\begin{eqnarray}
&~& M^2_{n,m}(\gamma, \gamma) = \pi^2 \left((n+\gamma)^2 + \frac{1}{3}(n+\gamma+2(m+\gamma))^2\right) 
\nonumber \\
&~& ~~~~~~~~~~~~~ = \frac{1}{3} \left[\left(\frac{n+\gamma}{R}\right)^2 
+ \left(\frac{n+\gamma}{R}\right)\left(\frac{m+\gamma}{R}\right) + \left(\frac{m+\gamma}{R}\right)^2\right] ,
\label{triplet-mass}
\end{eqnarray}
where the physical size $|e_1| = |e_2| = 2 \pi R$ is used in the final expression.
There are no massless modes from the $Z_3$ triplet in the case that $\gamma$ is not an integer.

\subsection{Classification of equivalence classes}

The classification of equivalence classes of BCs is reduced to the
classification of $(\Theta_0, \Theta_1)$.
We classify equivalence classes, which contain a set of diagonal representation matrices
$(\Theta_0, \Theta_1)$.
The diagonal matrices $(\Theta_0, \Theta_1)$ are specified by nine non-negative integers
$(l_p, m_p, n_p,$ $l_q, m_q, n_q, l_r, m_r, n_r)$ such that
\begin{eqnarray}
&~& \Theta_0 = {\mbox{diag}}(\overbrace{1, \cdots,~ 1, 1, \cdots,~ 1, 1, \cdots, 1}^p,
\overbrace{\omega, \cdots, \omega, \omega, \cdots, \omega, \omega, \cdots, \omega}^q
\nonumber \\
&~& ~~~~~~~~~~~~~~~~~~~~~~~~~~~~~~~~~~~~~~~~~~~~~~~~~~ \overbrace{\bar{\omega}, \cdots, \bar{\omega}, \bar{\omega}, 
\cdots, \bar{\omega}, \bar{\omega}, \cdots, \bar{\omega}}^{r=N-p-q}) , 
\nonumber \\
&~& \Theta_1 = {\mbox{diag}}(1, \cdots, 1, \omega, \cdots, \omega, \bar{\omega}, \cdots, \bar{\omega},
1, \cdots, 1, \omega, \cdots, \omega, \bar{\omega}, \cdots, \bar{\omega},
\nonumber \\
&~& ~~~~~~~~~~~~~~~~~~~~~~~~~~~~~~~~~~~~~~~~~~~~~~~~~~~ 1, \cdots, 1, \omega, \cdots, \omega, \bar{\omega}, \cdots, \bar{\omega}) ,
\nonumber \\
&~& \Theta_2 = {\mbox{diag}}(\underbrace{1, \cdots, 1}_{l_p}, \underbrace{\bar{\omega}, \cdots, \bar{\omega}}_{m_p} ,
\underbrace{\omega, \cdots, \omega}_{n_p}, \underbrace{\bar{\omega}, \cdots, \bar{\omega}}_{l_q},
\underbrace{\omega, \cdots, \omega}_{m_q}, \underbrace{1, \cdots, 1}_{n_q},
\nonumber \\
&~& ~~~~~~~~~~~~~~~~~~~~~~~~~~~~~~~~~~~~~~~~~~~~~~~~~~ 
\underbrace{\omega, \cdots, \omega}_{l_r}, \underbrace{1, \cdots, 1}_{m_r},
\underbrace{\bar{\omega}, \cdots, \bar{\omega}}_{n_r}) ,
\label{pqr}
\end{eqnarray}
where $\Theta_2$ is denoted, for convenience,
$N \geq l_p, m_p, n_p, l_q, m_q, n_q, l_r, m_r, n_r \geq 0$, $p=l_p + m_p + n_p$,
$q=l_q + m_q + n_q$ and $r=l_r + m_r + n_r$.
We denote each BC specified by $(l_p, m_p, n_p, l_q, m_q, n_q, l_r, m_r, n_r)$  (or a theory with such BCs)
as $[l_p, m_p, n_p;$ $l_q, m_q, n_q;$ $l_r, m_r, n_r]$.

The matrix $\Theta_1$ is interchanged with $\Theta_2$ by the following interchange among entries such that
\begin{eqnarray}
[l_p, m_p, n_p; l_q, m_q, n_q; l_r, m_r, n_r] 
\leftrightarrow [l_p, n_p, m_p; n_q, m_q, l_q; m_r, l_r, n_r]  .
\label{interchange12}
\end{eqnarray}
The matrix $\Theta_0$ is interchanged with $\Theta_1$ by the following interchange among entries such that
\begin{eqnarray}
[l_p, m_p, n_p; l_q, m_q, n_q; l_r, m_r, n_r] 
\leftrightarrow [l_p, l_q, l_r; m_p, m_q, m_r; n_p, n_q, n_r]  .
\label{interchange01}
\end{eqnarray}
The matrix $\Theta_0$ is interchanged with $\Theta_2$ by the following interchange among entries such that
\begin{eqnarray}
[l_p, m_p, n_p; l_q, m_q, n_q; l_r, m_r, n_r] 
\leftrightarrow [l_p, m_r, n_q; l_r, m_q, n_p; l_q, m_p, n_r]  .
\label{interchange02}
\end{eqnarray}

Using the equivalence relations (\ref{equX1}) and (\ref{equX2}), 
we can derive the following equivalence relations in
the $SU(N)$ gauge theory:
\begin{eqnarray}
&~& [l_p, m_p, n_p; l_q, m_q, n_q; l_r, m_r, n_r]  
\nonumber \\
&~&\sim [l_p-1, m_p+1, n_p; l_q, m_q-1, n_q+1; l_r+1, m_r, n_r-1] , 
\nonumber \\
&~& ~~~~~~~~~~~~~~~~~~~~~~~~~~~~~~~~~~~~~~~~~~~~~~~~~~~~~~~~~~~~~~~~~~ \mbox{for} ~~ l_p, m_q, n_r  \geq 1 ,
\nonumber \\
&~&\sim [l_p+1, m_p-1, n_p; l_q, m_q+1, n_q-1; l_r-1, m_r, n_r+1] , 
\nonumber \\
&~& ~~~~~~~~~~~~~~~~~~~~~~~~~~~~~~~~~~~~~~~~~~~~~~~~~~~~~~~~~~~~~~~~~~ \mbox{for} ~~ m_p, n_q, l_r  \geq 1 ,
\nonumber \\
&~&\sim [l_p-1, m_p, n_p+1; l_q+1, m_q-1, n_q; l_r, m_r+1, n_r-1] , 
\nonumber \\
&~& ~~~~~~~~~~~~~~~~~~~~~~~~~~~~~~~~~~~~~~~~~~~~~~~~~~~~~~~~~~~~~~~~~~ \mbox{for} ~~ l_p, m_q, n_r  \geq 1 ,
\nonumber \\
&~&\sim [l_p+1, m_p, n_p-1; l_q-1, m_q+1, n_q; l_r, m_r-1, n_r+1] , 
\nonumber \\
&~& ~~~~~~~~~~~~~~~~~~~~~~~~~~~~~~~~~~~~~~~~~~~~~~~~~~~~~~~~~~~~~~~~~~ \mbox{for} ~~ m_p, n_q, l_r  \geq 1 ,
\nonumber \\
&~&\sim [l_p, m_p-1, n_p+1; l_q+1, m_q, n_q-1; l_r-1, m_r+1, n_r] , 
\nonumber \\
&~& ~~~~~~~~~~~~~~~~~~~~~~~~~~~~~~~~~~~~~~~~~~~~~~~~~~~~~~~~~~~~~~~~~~ \mbox{for} ~~ m_p, n_q, l_r  \geq 1 ,
\nonumber \\
&~&\sim [l_p, m_p+1, n_p-1; l_q-1, m_q, n_q+1; l_r+1, m_r-1, n_r] , 
\nonumber \\
&~& ~~~~~~~~~~~~~~~~~~~~~~~~~~~~~~~~~~~~~~~~~~~~~~~~~~~~~~~~~~~~~~~~~~ \mbox{for} ~~ n_p, l_q, m_r  \geq 1 .
\label{equ-rel}
\end{eqnarray}
One can show that the number of equivalence classes of BCs including diagonal representations is 
$\displaystyle{{}_{N+8}C_8-2\cdot{}_{N+5}C_8}$ for the $SU(N)$ gauge group.

\subsection{Effective potential}

We study the effective potential for extradimensional components of the gauge field 
in an $SU(3)$ gauge theory on $M^4 \times (T^2/Z_3)$.
Let us adopt the representation matrices such that
\begin{eqnarray}
\Theta_0 = \Theta_1 = \Theta_2 =  
\left(
\begin{array}{ccc}
1 & 0 & 0 \\
0 & \omega & 0 \\
0 & 0 & \bar{\omega} 
\end{array}
\right) .
\label{Theta-ex}
\end{eqnarray}
\begin{wraptable}{l}{\halftext}
\caption{$(\theta_0, \theta_1, \theta_2)$ for gauge fields.}
\begin{center}
\begin{tabular}{c|ccc}\hline\hline
 & $\theta_0$ & $\theta_1$ & $\theta_2$ \\ \hline
$A_{\mu}^{1+}$, $A_{\mu}^{4-}$, $A_{\mu}^{6+}$ & $\bar{\omega}$ & $\bar{\omega}$ & $\bar{\omega}$ \\
$A_{\mu}^{1-}$, $A_{\mu}^{4+}$, $A_{\mu}^{6-}$ & $\omega$ & $\omega$ & $\omega$ \\
$A_{\mu}^3$, $A_{\mu}^8$ & 1 & 1 & 1 \\ \hline
$A_{z}^{1+}$, $A_{z}^{4-}$, $A_{z}^{6+}$ & $\omega$ & $\omega$ & $\omega$ \\
$A_{z}^{1-}$, $A_{z}^{4+}$, $A_{z}^{6-}$ & 1 & 1 & 1 \\
$A_{z}^3$, $A_z^8$ & $\bar{\omega}$ & $\bar{\omega}$ & $\bar{\omega}$ \\ \hline
$A_{\bar{z}}^{1+}$, $A_{\bar{z}}^{4-}$, $A_{\bar{z}}^{6+}$ & 1 & 1 & 1 \\
$A_{\bar{z}}^{1-}$, $A_{\bar{z}}^{4+}$, $A_{\bar{z}}^{6-}$ & $\bar{\omega}$ & $\bar{\omega}$ & $\bar{\omega}$ \\
$A_{\bar{z}}^3$, $A_{\bar{z}}^8$ & $\omega$ & $\omega$ & $\omega$ \\ \hline
\end{tabular}
\end{center}
\end{wraptable}
With this assignment, the eigenvalues $(\theta_0, \theta_1, \theta_2)$ for gauge fields are determined 
from the transformation properties under $Z_3$ transformation (\ref{A-z0}) -- (\ref{A-z2}),
and are given in Table I.
Here $(A_M^{1+}$, $\cdots$, $A_M^{6-})$ are defined by
\begin{eqnarray}
&~& A_M^{1+} \equiv \frac{1}{\sqrt{2}}\left(A_M^1 - i A_M^2\right) , \\
&~& A_M^{1-} \equiv \frac{1}{\sqrt{2}}\left(A_M^1 + i A_M^2\right) , \\
&~& A_M^{4+} \equiv \frac{1}{\sqrt{2}}\left(A_M^4 - i A_M^5\right) , \\
&~& A_M^{4-} \equiv \frac{1}{\sqrt{2}}\left(A_M^4 + i A_M^5\right) , \\
&~& A_M^{6+} \equiv \frac{1}{\sqrt{2}}\left(A_M^6 - i A_M^7\right) , \\
&~& A_M^{6-} \equiv \frac{1}{\sqrt{2}}\left(A_M^6 + i A_M^7\right) .
\label{A+-}
\end{eqnarray}
We find that zero modes appear in $A_{\mu}^3$, $A_{\mu}^8$, 
$A_{z}^{1-}$, $A_{z}^{4+}$, $A_{z}^{6-}$, $A_{\bar{z}}^{1+}$, $A_{\bar{z}}^{4-}$ and $A_{\bar{z}}^{6+}$.
{}From the vanishing field strength condition,
$A_z(x)$ and $A_{\bar{z}}(x)$ are parametrized as
\begin{eqnarray}
A_z(x) = \frac{\sqrt{2}\pi a}{g}   
\left(
\begin{array}{ccc}
0 & 0 & 1 \\
1 & 0 & 0 \\
0 & 1 & 0
\end{array}
\right) , ~~
A_{\bar{z}}(x) = \frac{\sqrt{2}\pi a}{g}   
\left(
\begin{array}{ccc}
0 & 1 & 0 \\
0 & 0 & 1 \\
1 & 0 & 0
\end{array}
\right) .
\label{Az-para}
\end{eqnarray}
Here, we set the zero mode $a$ to be real using the residual $U(1)$ gauge symmetries. 

Now we consider the effective potential $V_{\rm eff}$ for $a$
by treating it as a background field.
The effective potential is derived by writing
$A_M = A^0_M + A^q_M$, taking a suitable gauge fixing and integrating over
the quantum part $A^q_M$ and every quantum flactuation of other fields.
Here $A^0_M$ is a background configuration of the gauge field $A_M$.
The $V_{\rm eff}$ depends not only on $A^0_M$ but
also on BCs, i.e., $V_{\rm eff} = V_{\rm eff} [A^0_M; \Theta_0, \Theta_1, \gamma]$.
If the gauge fixing term is also invariant under the gauge
transformation, i.e.,
\begin{eqnarray}
D^{M}(A^{0}) A_M = 0 \to D^{M}({A'}^{0}) A'_M = \Omega D^{M}(A^{0}) A_M
\Omega^{{\dagger}} = 0 ,
\label{gauge-inv-gf}
\end{eqnarray}
it is shown that $V_{\rm eff}$ satisfies
\begin{eqnarray}
V_{\rm eff} [A^0_M; \Theta_0, \Theta_1, \gamma] = V_{\rm eff} [A'^0_M; \Theta'_0, \Theta'_1, \gamma] .
\label{Veff}
\end{eqnarray}
This property implies that the minimum $V_{\rm eff}$ corresponds to the same
symmetry as that of $(\Theta_0^{\rm sym}, \Theta_1^{\rm sym})$ .

The one-loop effective potential is given by
\begin{eqnarray}
V_{\rm eff} [A^0_M; \Theta_0, \Theta_1, \gamma] &=& \sum \mp {i \over 2} {\mbox{Tr}}~{\mbox{ln}}
D_M(A^0) D^M(A^0)  ,
\label{Veff1}\\
 &=& \sum \mp {1 \over 2} \int {d^4 p_E \over (2 \pi)^4} \sum_{n, m} \ln (p_E^2 +  \hat{M}_{n, m}^2 - i \varepsilon) ,
\label{Veff2}
\end{eqnarray}
where $p_E$ is a four-dimensional Euclidean momentum and the Wick rotation is applied.  
Here, we consider that $F_{MN}^0 = 0$ and every field has no mass term on six-dimensional space-time.
The sums extend over all degrees of freedom of fields in the bulk in Eq. (\ref{Veff1})
and over all degrees of freedom of four-dimensional fields whose masses are $\hat{M}_{n, m}$ in Eq. (\ref{Veff2}).
The sign is negative (positive) for bosons (FP ghosts and fermions).
$D_M(A^0)$ denotes an appropriate covariant derivative with respect to $A^0_M$. 
For later convenience, we write down the formula of one-loop effective potential
for $\hat{M}_{n, m}^2 = M_{n, m}^2(\alpha, \beta)$ as\cite{AB&Q} 
\begin{eqnarray}
&~& V_{\rm eff} [A^0_M; \Theta_0, \Theta_1, \gamma] = \sum_{(\alpha, \beta)} \mp {1 \over 2} I(\alpha, \beta) ,
\nonumber \\
&~& I(\alpha, \beta) \equiv \int {d^4 p_E \over (2 \pi)^4} \sum_{n, m} 
\ln (p_E^2 +  {M}_{n, m}^2(\alpha, \beta) - i \varepsilon) 
\nonumber \\
&~& ~~~~~~~~~ =  \frac{\sqrt{3}}{256\pi^7R^4}{\sum_{n,m}}' \frac{1}{(n^2+m^2-nm)^3}
\cos 2\pi (\alpha n+ \beta m) 
\nonumber \\
&~& ~~~~~~~~~~~~~~~~~~ + (\alpha, \beta \mbox{-independent terms}) .
\label{Veff3}
\end{eqnarray}

\begin{wraptable}{l}{\halftext}
\caption{$(\theta_0, \theta_1, \theta_2)$ for $\phi$.}
\begin{center}
\begin{tabular}{c|ccc}\hline\hline
 & $\theta_0$ & $\theta_1$ & $\theta_2$ \\ \hline
$\phi^1$ & 1 & 1 & 1 \\
$\phi^2$ & $\omega$ & $\omega$ & $\omega$ \\
$\phi^3$ & $\bar{\omega}$ & $\bar{\omega}$ & $\bar{\omega}$ \\ \hline
\end{tabular}
\end{center}
\end{wraptable}
Our task now is to obtain mass squareds $\hat{M}_{n, m}^2$ for every field that couples to gauge fields.
For simplicity, we consider an $SU(3)$ triplet scalar field $\phi = (\phi^1, \phi^2, \phi^3)$
whose eigenvalues $(\theta_0, \theta_1, \theta_2)$ are given in Table II.
The point is to consider $\phi$ as a $Z_3$ triplet, 
i.e., $\phi_{n,m}^1(x) = \phi_{n,m}^2(x) = \phi_{n,m}^3(x) \equiv \phi_{n,m}(x)$.
Then the covariant derivative for $\phi(x, z, \bar{z})$ is calculated as
\begin{eqnarray}
&~&D_z \phi = (\partial_z + i g A_z) \phi 
= {\sum_{n,m}} \phi_{n,m}(x) 
\left(
\begin{array}{ccc}
\partial_z & 0 & i \sqrt{2}\pi a  \\
i \sqrt{2}\pi a & \partial_z & 0  \\
0 & i \sqrt{2}\pi a  & \partial_z
\end{array}
\right)  
\left(
\begin{array}{c}
f_{n,m}^{(0)} \\
f_{n,m}^{(1)} \\
f_{n,m}^{(2)} 
\end{array}
\right) 
\nonumber \\
&~& = {\sum_{n,m}} \phi_{n,m}(x) 
\left(
\begin{array}{c}
i\pi \left(n-\frac{a}{\sqrt{2}} + \frac{n-\frac{a}{\sqrt{2}} 
 + 2\left(m-\frac{a}{\sqrt{2}}\right)}{\sqrt{3}}i\right)
\bar{\omega}f_{n,m}^{(2)} \\
i\pi \left(n-\frac{a}{\sqrt{2}} + \frac{n-\frac{a}{\sqrt{2}} 
 + 2\left(m-\frac{a}{\sqrt{2}}\right)}{\sqrt{3}}i\right)
\bar{\omega}f_{n,m}^{(0)} \\
i\pi \left(n-\frac{a}{\sqrt{2}} + \frac{n-\frac{a}{\sqrt{2}} 
 + 2\left(m-\frac{a}{\sqrt{2}}\right)}{\sqrt{3}}i\right)
\bar{\omega}f_{n,m}^{(1)} 
\end{array}
\right) .
\label{Dzphi}
\end{eqnarray}
Hence the mass squareds for $\phi_{n,m}(x)$ are three $M_{n,m}^2(-\frac{a}{\sqrt{2}}, -\frac{a}{\sqrt{2}})$'s.
In the same way, those of gauge fields are calculated from the covariant derivative 
$D_z A_M = \partial_z A_M + ig[A_z, A_M]$ and 
are $M_{n,m}^2(-a, -a)$, $M_{n,m}^2(\frac{1+\sqrt{3}}{2}a, \frac{1+\sqrt{3}}{2}a)$, 
$M_{n,m}^2(\frac{1-\sqrt{3}}{2}a, \frac{1-\sqrt{3}}{2}a)$  
and five $M_{n,m}^2(0, 0)$'s.
The same result holds for FP ghosts.

Using mass squareds and Eq. (\ref{Veff3}), we obtain the one-loop effective potential for $a$ as
\begin{eqnarray}
&~& V_{\rm eff} = - 2 I(-a, -a) - 2 I\left(\frac{1+\sqrt{3}}{2}a, \frac{1+\sqrt{3}}{2}a\right)
- 2 I\left(\frac{1-\sqrt{3}}{2}a, \frac{1-\sqrt{3}}{2}a\right) 
\nonumber \\
&~& ~~~~~~~~ - \frac{3}{2} I\left(-\frac{a}{\sqrt{2}},-\frac{a}{\sqrt{2}}\right) .
\label{V}
\end{eqnarray}
The minimum $V_{\rm eff}$ is given at $a = 0$.
When fermions are introduced, the non-vanishing expectation value of $a$ can be obtained and 
the breakdown of $U(1)$ gauge symmetries can occur.

\section{Conclusions}

We have studied equivalence classes of BCs in a gauge theory on the orbifold $T^2/Z_3$.
General arguments have been given for BCs in gauge theories on $T^2/Z_3$ 
including various relations of BCs, and
equivalence classes of BCs have been defined by the invariance under gauge transformation.
Mode expansions have been given for six-dimensional $Z_3$ singlet fields and the $Z_3$ triplet field,
and the classification of BCs for the $SU(N)$ gauge group has been carried out
with the aid of equivalence relations.
The one-loop effective potential for Wilson line phases has been calculated
using the $SU(3)$ gauge theory.
It is crucial to study dynamical gauge symmetry breaking and mass generation
in a realistic model including fermions.
It is also important to construct a phenomenologically viable model 
realizing gauge-Higgs unification\cite{GHU} and/or family unification\cite{OFU} based on them.
The local grand unification can be realized by taking nontrivial $\Theta_i$'s.\footnote{
The $\lq$local' gauge groups at fixed points were realized on $T^2/Z_2$ in Ref.~\citen{ABC}.
The string-derived orbifold grand unification theories were studied in Refs.~\citen{KRZ} and \citen{BHLR}.}
It is interesting to study the phenomenological aspects of such models.
We hope to further study these subjects in the near future.

\section*{Acknowledgements}
This work was supported in part by Scientific Grants from the Ministry of Education, Culture, Sports and Technology
under Grant Nos.18204024 and 18540259 (Y.~K.).

\appendix
\section{Useful Formulae}

For $Y_+^k$, $Y_-^k$ $(k=1,2,3)$ and $X$ defined by
\begin{eqnarray}
\hspace{-0.0cm}&~& Y_+^1 = 
\left(
\begin{array}{ccc}
0 & 1 & 0 \\
0 & 0 & 1 \\
1 & 0 & 0 
\end{array}
\right), ~~
Y_+^2 = 
\left(
\begin{array}{ccc}
0 & \omega & 0 \\
0 & 0 & \omega \\
\omega & 0 & 0 
\end{array}
\right), ~~
Y_+^3 = 
\left(
\begin{array}{ccc}
0 & \omega^2 & 0 \\
0 & 0 & \omega^2 \\
\omega^2 & 0 & 0 
\end{array}
\right) , 
\nonumber\\
\hspace{-0.0cm}&~& Y_-^1 = 
\left(
\begin{array}{ccc}
0 & 0 & 1 \\
1 & 0 & 0 \\
0 & 1 & 0 
\end{array}
\right), ~~
Y_-^2 = 
\left(
\begin{array}{ccc}
0 & 0 & \omega^2 \\
\omega^2 & 0 & 0 \\
0 & \omega^2 & 0 
\end{array}
\right), ~~
Y_-^3 = 
\left(
\begin{array}{ccc}
0 & 0 & \omega \\
\omega & 0 & 0 \\
0 & \omega & 0 
\end{array}
\right) ,
\nonumber \\
\hspace{-0.0cm}&~& X = 
\left(
\begin{array}{ccc}
1 & 0 & 0 \\
0 & \omega & 0 \\
0 & 0 & \omega^2 
\end{array}
\right),  
\label{Y+-X}
\end{eqnarray}
the following relation holds:
\begin{eqnarray}
X\exp\left[-i\sum_{k=1}^3 \left(a^k Y_+^k + \bar{a}^k Y_-^k\right)\right]
 = \exp\left[-i\sum_{k=1}^3 \left(\bar{\omega} a^k Y_+^k + \omega \bar{a}^k Y_-^k\right)\right]X .
\label{XYrelation}
\end{eqnarray}

For $Y$, $Y_{\omega}$ and $Y_{\bar{\omega}}$ defined by
\begin{eqnarray}
\hspace{-0.7cm} Y = 
\left(
\begin{array}{ccc}
0 & 1 & 1 \\
1 & 0 & 1 \\
1 & 1 & 0 
\end{array}
\right), ~~
Y_{\omega} = 
\left(
\begin{array}{ccc}
0 & \omega & \omega^2 \\
\omega^2 & 0 & \omega \\
\omega & \omega^2 & 0 
\end{array}
\right), ~~
Y_{\bar{\omega}} = 
\left(
\begin{array}{ccc}
0 & \omega^2 & \omega \\
\omega & 0 & \omega^2 \\
\omega^2 & \omega & 0 
\end{array}
\right) , 
\label{Ys2}
\end{eqnarray}
the $n$-th powers of $Y$, $Y_{\omega}$ and $Y_{\bar{\omega}}$ are calculated as
\begin{eqnarray}
&~& Y^n = \frac{1}{3}\left(2^n - (-1)^n\right) Y + \frac{1}{3}\left(2^n + 2 (-1)^n\right) I ,
\nonumber \\
&~& Y_{\omega}^n = \frac{1}{3}\left(2^n - (-1)^n\right) Y_{\omega} + \frac{1}{3}\left(2^n + 2 (-1)^n\right) I ,
\nonumber \\
&~& Y_{\bar{\omega}}^n = \frac{1}{3}\left(2^n - (-1)^n\right) Y_{\bar{\omega}} + \frac{1}{3}\left(2^n + 2 (-1)^n\right) I .
\label{Yn}
\end{eqnarray} 
Then $e^{iaY}$, $e^{iaY_{\omega}}$ and $e^{iaY_{\bar{\omega}}}$ are calculated as
\begin{eqnarray}
&~& e^{iaY} = \frac{1}{3}\left(e^{2ai} - e^{-ai}\right) Y + \frac{1}{3}\left(e^{2ai} + 2 e^{-ai}\right) I ,
\nonumber \\
&~& e^{iaY_{\omega}} = \frac{1}{3}\left(e^{2ai} - e^{-ai}\right) Y_{\omega} + \frac{1}{3}\left(e^{2ai} + 2 e^{-ai}\right) I ,
\nonumber \\
&~& e^{iaY_{\bar{\omega}}} = \frac{1}{3}\left(e^{2ai} - e^{-ai}\right) Y_{\bar{\omega}} 
+ \frac{1}{3}\left(e^{2ai} + 2 e^{-ai}\right) I .
\label{expY}
\end{eqnarray}

For the function $f_{n+\alpha,m+\beta}(z, \bar{z})$ defined by
\begin{eqnarray}
&~& f_{n+\alpha,m+\beta}(z, \bar{z})  
\equiv \exp \left[\pi i\left\{\left(n + \alpha - \frac{n + \alpha + 2(m + \beta)}{\sqrt{3}}i \right) z\right.\right.
\nonumber \\
&~& ~~~~~~~~~~~~~~~~~~~~~~~~~~~~~~~ \left.\left. + \left(n + \alpha + 
\frac{n + \alpha + 2(m + \beta)}{\sqrt{3}}i \right) \bar{z} \right\}\right] , 
\label{f2}
\end{eqnarray}
the following transformation properties are derived:
\begin{eqnarray}
&~& f_{n+\alpha,m+\beta}(z+1, \bar{z}+1) = \omega^{3\alpha} f_{n+\alpha,m+\beta}(z, \bar{z}) , 
\nonumber \\
&~& f_{n+\alpha,m+\beta}(z+\omega, \bar{z}+\bar{\omega}) = \omega^{3\beta} f_{n+\alpha,m+\beta}(z, \bar{z}) ,
\nonumber \\ 
&~& f_{n+\alpha,m+\beta}(z+\bar{\omega}, \bar{z}+{\omega}) = \bar{\omega}^{3(\alpha+\beta)} f_{n+\alpha,m+\beta}(z, \bar{z}) ,
\nonumber \\ 
&~& f_{n+\alpha,m+\beta}(\omega z, \bar{\omega}\bar{z}) = f_{m+\beta, -m-n-\alpha-\beta}(z, \bar{z}) , 
\nonumber \\
&~& f_{n+\alpha,m+\beta}(\bar{\omega} z, \omega\bar{z}) = f_{-m-n-\alpha-\beta, n+\alpha}(z, \bar{z}) .
\label{ftran}
\end{eqnarray}


\begin{thebibliography}{99}

\bibitem{H}
Y.~Hosotani, \PLB{126,1983,309}; \ANN{190,1989,233}.

\bibitem{Kawamura1}
Y.~Kawamura, \PTP{103,2000,613}; ibid. \andvol{105,2001,999}.

\bibitem{Hall1}
L.~Hall and Y.~Nomura, \PRD{64,2001,055003}.

\bibitem{Higgs}
E.~Witten, \NPB{268,1985,75};

L.~E.~Ib\'a\~nez, J.~E.~Kim, H.~P.~Nilles and F.~Quevedo, \PLB{191,1987,282}.

\bibitem{HHHK}
N.~Haba, M.~Harada, Y.~Hosotani and Y.~Kawamura.
\NPB{657,2003,169} [Errata; B \textbf{669} (2003), 381].

\bibitem{HHK}
N.~Haba, Y.~Hosotani and Y.~Kawamura, \PTP{111,2004,265}.     

\bibitem{KLY}
M.~Kubo, C.~S.~Lim and H.~Yamashita, Mod.~Phys.~Lett. A \textbf{17} (2002), 2249.

\bibitem{HN&T}
Y.~Hosotani, S.~Noda and K.~Takenaga, \PRD{69,2004,125014}.

\bibitem{threefamilies}
T.~Watari and T.~Yanagida, \PLB{532,2002,252}.

K.~S.~Babu, S.~M.~Barr and B.~Kyae, \PRD{65,2002,115008}.

\bibitem{GLM&S}
I.~Gogoladze, C.~A.~Lee, Y.~Mimura and Q.~Shafi, \PLB{649,2007,212}.

\bibitem{orbifold}
L.~Dixon, J.~Harvey, C.~Vafa and E.~Witten, \NPB{261,1985,678}; \NPB{274,1986,285}.

\bibitem{SS}
J.~Scherk and J.~H.~Schwarz, \PLB{82,1979,60}; \NPB{153,1979,61}.

\bibitem{HKO}
N.~Haba, Y.~Kawamura and K. Oda, arXiv:0803.4380.

\bibitem{AB&Q}
I.~Antoniadis, K.~Benakli and M.~Quiros, New J. Phys. \textbf{3} (2001), 20.1.

\bibitem{GHU}
L.~J.~Hall, Y.~Nomura and D.~T.-Smith, \NPB{639,2002,307}.

N. Haba, Y. Hosotani, Y. Kawamura and T. Yamashita, \PRD{70,2004,015010}. 

\bibitem{OFU}
Y. Kawamura, T. Kinami and K. Oda, \PRD{76,2007,035001}.

Y. Kawamura and T. Kinami, Int. J. Mod. Phys. A \textbf{22} (2007), 4617; \PTP{119,2008,285}. 

\bibitem{ABC}
T.~Asaka, W.~Buchm\"uller and  L.~Covi, \PLB{523,2001,199}; \PLB{540,2002,295}.

\bibitem{KRZ}
T.~Kobayashi, S.~Raby and R.-J.~Zhang, \PLB{593,2004,262}.

\bibitem{BHLR}
W. Buchm\"uller, K. Hamaguchi, O. Lebedev and M. Ratz, \PRL{96,2006,121602}; \NPB{785,2007,149}.

\end{thebibliography}
\end{document}